# Explanation of multicollinearity using the decomposition theorem of ordinary linear regression models[1]

Xingguo Wu

(DongGuan Polytechnic, Dongguan, Guangdong, 523808, China)

**Abstract**

In a multiple linear regression model, the algebraic formula of the decomposition theorem explains the relationship between the univariate regression coefficient and partial regression coefficient using geometry. It was found that univariate regression coefficients are decomposed into their respective partial regression coefficients according to the parallelogram rule. Multicollinearity is analyzed with the help of the decomposition theorem. It was also shown that it is a sample phenomenon that the partial regression coefficients of important explanatory variables are not significant, but the sign expectation deviation cause may be the population structure between the explained variables and explanatory variables or may be the result of sample selection. At present, some methods of diagnostic multicollinearity only consider the correlation of explanatory variables, so these methods are basically unreliable, and handling multicollinearity is blind before the causes are not distinguished. The increase in the sample size can help identify the causes of multicollinearity, and the difference method can play an auxiliary role.

**Keywords:** multicollinearity; decomposition theorem; difference structure invariance theorem; difference method; geometric interpretation

## 1. Introduction

In conventional linear regression models, the non-existence of complete multicollinearity between selected explanatory variables was required. In general, complete multicollinearity

---

[1] Table of Abbreviations

FWL           Frisch-Waugh-Lovell

can be solved well. However, in reality, a certain correlation between the explanatory variables is almost inevitable, and many classic textbooks define this correlation as multicollinearity. According to this definition, the existence of multicollinearity is not as important in comparison with its degree of strength. However, the term multicollinearity can also be differently interpreted, and some textbooks define multicollinearity as follows. If two or more explanatory variables are highly correlated, it is difficult to distinguish the influences and effects of one explanatory variable on an explained variable. The consequence of the correlations between explanatory variables is defined as multicollinearity. This paper used the latter definition to explain the problem of multicollinearity.

Multicollinearity was first proposed by Frisch in 1934. Since then, many researchers have participated in in-depth discussions regarding this issue, and two different attitudes were formed, which Fanar and Glauber have previously summarized (1967). Some believe that multicollinearity does not require handling because it does not affect the BLUE nature of OLS estimators, such as Mason et al. (1991). Others believe that the consequences of multicollinearity cannot be ignored and that they need to be corrected. However, there are many conflicts concerning the understanding of multicollinearity. (1) Concerning the cause of multicollinearity, some econometricians believe that it is caused by the nature of variables, such as the existence of too many explanatory variables, inappropriate regression equation forms, etc. Nevertheless, some scholars believe that multicollinearity is a sample phenomenon. For example, Blanchard thinks that multicollinearity is a problem of insufficient sample data (Blanchard (1967) and Goldberger (1991)). Although these reasons may lead to multicollinearity, they do not essentially reflect its problem. (2) There are some methods for diagnosing multicollinearity, such as examining the partial correlation coefficients, tolerances, variance inflation factors, conditional indices, etc. However, none of these detection methods are generally accepted. For example, C. Robert (1975) constructed an example illustrating the failure of the partial correlation test. (3) The potential solutions to solving multicollinearity are as follows. Bowerman and O'Connell (1993) and Bring (1996) stated the issues related to eliminating problematic variables from the model, Burt (1987) considered the application of the difference method, and Hoerl and Kennard (1970) proposed a ridge estimation method, where Dijkstra (2014) and Mansson et al. (2014) made further research on it. Also, Silvey (1969) and Gower and Blasius (2005) introduced principal component analytical methods to address the multicollinearity problem, and Walker (1989) developed a diagnostic tool for detecting the observations of the effects of collinearity. Each method has its limitations, and the blind use of these methods has also attracted severe criticism from their peers.

So far, there is still much controversy regarding the interpretation of multicollinearity and its causes, diagnosis, and treatment. Also, there have not been any substantial breakthroughs. The purpose of this article is to try to provide a reasonable geometric interpretation of multicollinearity. The method of solving multicollinearity in regression models needs to be chosen by researchers according to their own needs. The rest of the paper is organized as follows. Section 2 presents a decomposition theorem existing in a derived regression model. Section 3, the decomposition theorem is stated geometrically. Section 4, the multicollinearity phenomenon is explained using the decomposition theorem. Section 5 explains a Monte Carlo simulation experiment. Section 6 analyzes the shortcomings of the existing methods in regards to dealing with multicollinearity, and Section 7 is the conclusion.

## 2. Decomposition theorem

In a general multiple linear regression model, the expression is as follows.

$$y = X * (\beta_0, \beta)^T + u \qquad (1)$$

The regression coefficient $(X^T X)^{-1} X^T y = (\hat{\beta}_0, \hat{\beta})^T$ is the optimal estimator based on the least-squares method, and the regression coefficient $\hat{\beta} = (\hat{\beta}_1, \hat{\beta}_2, \cdots, \hat{\beta}_p) \in R^p$ is a linear combination of the explained variable $y$. It is well known that the intercept $\hat{\beta}_0$ plays a translational role. The meaning of the regression coefficient $\hat{\beta}$ we are concerned with can be expressed as follows. When an explanatory variable $x$ changes by one unit, the explained variable $y$ changes by $\hat{\beta}$ units on average since its economic meaning is expressed as the average change in $y$. Then, it is expressed as a linear combination of $\triangle y$. The modern econometric theory pays too much attention to it as a linear combination of $y$ and ignores that it is also a linear combination of $\triangle y$. In univariate linear regression, $\hat{\beta}_n$ can be rewritten as:

$$\hat{\beta}_n = w_{(1)} \triangle y_{(1)} + w_{(2)} \triangle y_{(2)} + \cdots + w_{(n-1)} \triangle y_{(n-1)}. \tag{2}$$

Here, the subscript indicates the serial number of the observation, $n$ is the sample size, $\bar{x}$ is the mean of the explanatory variable $x$, $s_x^2$ is the sample variance, and $g_{(i)} = (\bar{x} - x_{(i)})/ns_x^2$ is defined. The cumulative sum $w_{(j)} = g_{(1)} + g_{(2)} + \cdots + g_{(j)}$ is the weight of $\triangle y_{(j)}$, where $\triangle y_{(j)}$ is defined as $\triangle y_{(j)} = y_{(j+1)} - y_{(j)}$. Since $(\bar{x} - x_{(1)}) + \cdots + (\bar{x} - x_{(n)}) = 0$, then $w_{(n)} = 0$. At the same time, $\triangle y_{(i)} = y_{(i+1)} - y_{(i)} = \triangle \hat{y}_{(i)} + \triangle \hat{u}_{(i)} = \hat{\beta}_{(i)} \triangle x_{(i)} + \triangle \hat{u}_{(i)}$ and $\triangle x_{(i)} = x_{(i+1)} - x_{(i)}$. Equation (2) can be rewritten as:

$$\begin{aligned}\hat{\beta}_n &= w_{(1)} \triangle \hat{y}_{(1)} + w_{(2)} \triangle \hat{y}_{(2)} + \cdots + w_{(n-1)} \triangle \hat{y}_{(n-1)} \\ &\quad + w_{(1)} \triangle \hat{u}_{(1)} + w_{(2)} \triangle \hat{u}_{(2)} + \cdots + w_{(n-1)} \triangle \hat{u}_{(n-1)} \\ &= w_{(1)} \hat{\beta}_{(1)} \triangle x_{(1)} + w_{(2)} \hat{\beta}_{(2)} \triangle x_{(2)} + \cdots + w_{(n-1)} \hat{\beta}_{(n-1)} \triangle x_{(n-1)} \\ &\quad + w_{(1)} \triangle \hat{u}_{(1)} + w_{(2)} \triangle \hat{u}_{(2)} + \cdots + w_{(n-1)} \triangle \hat{u}_{(n-1)} \\ &= \hat{\beta}(w_{(1)} \triangle x_{(1)} + w_{(2)} \triangle x_{(2)} + \cdots + w_{(n-1)} \triangle x_{(n-1)})\end{aligned}$$

（if $\hat{\beta}_{(1)} = \hat{\beta}_{(2)} = \cdots = \hat{\beta}_{(n-1)} = \hat{\beta}$）

Then, it is easy to verify that $\sum w_{(i)} \triangle x_{(i)} = 1, \sum w_{(i)} \triangle \hat{u}_{(i)} = 0$。

**Lemma 1**. In univariate linear regression, the estimated coefficient $\hat{\beta}$ of $y$ on $x$ can be expressed as the inner product of the cumulative sum vector $w$ of $x$ and the difference vector $\triangle y$ or $\triangle \hat{y}$ of $y$, namely:

$$[w, \triangle y] = [w, \triangle \hat{y}] = \hat{\beta} \tag{3}$$

Here, $w = (w_{(1)}, w_{(2)}, \ldots, w_{(n-1)})$ and $\triangle y = (\triangle y_{(1)}, \triangle y_{(2)}, \ldots, \triangle y_{(n-1)})$ can be defined as above. To extend it further, if we make the inner product of the cumulative sum vector $w$ and the difference vector $\triangle x_j$ of $x_j$, his inner product form becomes $[w, \triangle x_j] = \hat{b}_{j \cdot i}$, where $\hat{b}_{j \cdot i}$ represents the regression coefficient of $x_j$ on $x_i$. Obviously, if there is the inner product of itself, then $[w, \triangle x_i] = \hat{b}_{i \cdot i} = 1$. If a small sample is generalized to a large sample, when $n \Rightarrow \infty$, then

$$\hat{\beta}_n = \sum_{j=1}^{n-1} \hat{\beta}_{(j)} w_{(j)} \triangle x_{(j)} = \frac{1}{n} \sum_{j=1}^{n-1} \hat{\beta}_{(j)} (\sum_{i=1}^{j} \frac{\overline{x} - x_{(i)}}{s_x^2}) \triangle x_{(j)}$$
$$\xrightarrow{n \to \infty} \int_{-\infty}^{+\infty} \beta_{(x)} \int_{-\infty}^{x} \frac{E(x) - t}{\sigma_x^2} dt dx = E(\beta_{(x)}) \tag{4}$$

where $\beta_{(x)}$ represents the regression coefficient under the given explanatory variable $x$. The expected value $E(\beta_{(x)})$ of the regression coefficient is the Epanechnikov kernel estimate of the regression coefficient $\beta_{(x)}$ with the kernel function $\int_{-\infty}^{x} \frac{E(x) - t}{\sigma_x^2} dt$ as the weight. It just takes the mean of the explanatory variable as the center and the entire sample area as the bandwidth.

In order to distinguish the regression symbol in the multiple linear regression model, we used $\dot{\beta}$ for the univariate regression coefficients and $\beta$ for the partial regression coefficients, where its form is:

$$\hat{y} = \hat{\beta}_0 + \hat{\beta}_1 x_1 + \hat{\beta}_2 x_2 + \cdots + \hat{\beta}_p x_p. \tag{5}$$

If the multiple regression model is stable, the difference form of equation (5) can be obtained:

$$\triangle \hat{y} = \hat{\beta}_1 \triangle x_1 + \hat{\beta}_2 \triangle x_2 + \cdots + \hat{\beta}_p \triangle x_p. \tag{6}$$

According to Lemma 1, the unary estimated parameter $\hat{\dot{\beta}}_1$ of $y$ on $x_1$ can be expressed as

$$\begin{aligned}\hat{\dot{\beta}}_1 &= [w, \triangle \hat{y}] \\ &= [w, \hat{\beta}_1 \triangle x_1 + \hat{\beta}_2 \triangle x_2 + \cdots + \hat{\beta}_p \triangle x_p] \\ &= \hat{\beta}_1 \hat{b}_{1 \cdot 1} + \hat{\beta}_2 \hat{b}_{2 \cdot 1} + \cdots + \hat{\beta}_p \hat{b}_{p \cdot 1} \\ &= \hat{\beta}_1 + \hat{\beta}_2 \hat{b}_{2 \cdot 1} + \cdots + \hat{\beta}_p \hat{b}_{p \cdot 1}\end{aligned}.$$

The unary regression coefficient $\hat{\beta}_1$ represents the total effect of $y$ on $x_1$, which is decomposed into the direct net effect $\hat{\beta}_1$ including itself and also into the indirect effect $\sum_{j\neq 1}^{p} \hat{\beta}_j \hat{b}_{j\cdot 1}$ through other explanatory variables. The univariate regression coefficients of the variables are decomposed into partial regression coefficients. For any explanatory variable $x_i$, the relationship between the univariate regression coefficient $\hat{\beta}_i$ and all the partial regression coefficient vectors [$\hat{\beta}_j, j = 1\cdots, p$] can be expressed as follows:

$$\hat{\hat{\beta}}_i = \sum_{j=1}^{p} \hat{\beta}_j \hat{b}_{j\cdot i} . \tag{7}$$

Here, $\hat{b}_{j\cdot i}$ represents the univariate regression coefficient of the explanatory variable $x_j$ on $x_i$. Obviously, if it is regressed on itself, then $\hat{b}_{i\cdot i} = 1$. Equation (7) can be represented by the following matrix:

$$\begin{pmatrix} \hat{\hat{\beta}}_1 \\ \hat{\hat{\beta}}_2 \\ \vdots \\ \hat{\hat{\beta}}_p \end{pmatrix} = B \begin{pmatrix} \hat{\beta}_1 \\ \hat{\beta}_2 \\ \vdots \\ \hat{\beta}_p \end{pmatrix} = \begin{pmatrix} \hat{b}_{1\cdot 1} & \hat{b}_{2\cdot 1} & \cdots & \hat{b}_{p\cdot 1} \\ \hat{b}_{1\cdot 2} & \hat{b}_{2\cdot 2} & \cdots & \hat{b}_{p\cdot 2} \\ \vdots & \vdots & \cdots & \vdots \\ \hat{b}_{1\cdot p} & \hat{b}_{2\cdot p} & \cdots & \hat{b}_{p\cdot p} \end{pmatrix} \begin{pmatrix} \hat{\beta}_1 \\ \hat{\beta}_2 \\ \vdots \\ \hat{\beta}_p \end{pmatrix} = \begin{pmatrix} 1 & \hat{b}_{2\cdot 1} & \cdots & \hat{b}_{p\cdot 1} \\ \hat{b}_{1\cdot 2} & 1 & \cdots & \hat{b}_{p\cdot 2} \\ \vdots & \vdots & \cdots & \vdots \\ \hat{b}_{1\cdot p} & \hat{b}_{2\cdot p} & \cdots & 1 \end{pmatrix} \begin{pmatrix} \hat{\beta}_1 \\ \hat{\beta}_2 \\ \vdots \\ \hat{\beta}_p \end{pmatrix} . \tag{8}$$

This is the decomposition theorem in multiple linear regression. Whether in a large sample or a small sample, this theorem is held. In a multiple linear regression model with $p$ explanatory variables, the unary linear regression coefficients of the $p$ explanatory variables are decomposed into partial regression coefficients according to the matrix $B$, which reflects the structural relationship between the explanatory variables. Adding or subtracting the explanatory variables is nothing more than redistributing the univariate linear regression coefficients of the explanatory variables according to the new structure. If the explanatory variables are independent of each other, then the matrix $B$ is the identity matrix, and the partial regression coefficient is the univariate regression coefficient. At present, statisticians and economists have a bias in understanding the consistency of estimated parameters. The former believe that any regression model is an approximation to the real complex world and that it is impossible to build a truly complete model. A model can only be explained based on the structure of the existing explanatory variables, and when a sample trends infinitely, the estimated parameters of OLS still have consistency. The structure of explanatory variables is a prerequisite for discussing the consistency of estimated parameters. If the structure of explanatory variables is different, then it is inappropriate to require the same consistency of the estimated parameters in models with different structures. The different structures of explanatory variables show that people explore the real world from different angles and that it is impossible to completely fit a truly complex world in a social environment. For economists, the premise that the specification of a model is assumed to be correct is particularly important. When adding redundant variables and missing relevant variables, model specification bias may cause inconsistent parameter estimates according to the decomposition theorem. However, they do not have a correct specification standard of the model, and there is no known prophet. Like many consumption theories, there is no standard answer to which specification of a theoretical model is correct. For a more detailed discussion of the decomposition theorem, you need to refer to Wu Xingguo and Lei Qinli (2020).

## 3. Geometric interpretation

In modern econometrics, algebra occupies a very important position. The definition of the concept and proof of the theory are basically based on algebra. Relatively speaking, geometric analyses are poor. However, geometry is a powerful tool in regression analysis, and geometric analyses are more helpful in recognizing the nature of things (Saville (1986)). In this section, we used geometry as a tool to state and explain the decomposition theorem. Before the analysis, we provided a general geometric definition. The multivariate linear regression model takes the form:

$$Y = \beta_0 + \beta X + u, \qquad (9)$$

where $n$ is the sample size. In geometry, the explained variables, explanatory variables, and unobserved error terms are all regarded as n-dimensional vectors. For example，$\vec{Y} = (Y_1, Y_2, \cdots, Y_n)$ is an n-dimensional observation vector, and the intercept $\beta_0 = \bar{Y} - \beta \bar{X}$ acts as a translational motion. We centralized all the variables as follows:

$$\vec{y} = \vec{Y} - \vec{\overline{Y}} = (Y_{(1)} - \overline{Y}, Y_{(2)} - \overline{Y}, \cdots, Y_{(n)} - \overline{Y}),$$

$$\vec{x}_i = \vec{X}_i - \vec{\overline{X}}_i = (X_{i(1)} - \overline{X}_i, X_{i(2)} - \overline{X}_i, \cdots, X_{i(n)} - \overline{X}_i).$$

The vector $\vec{y}$ is a span in the vector space $(\vec{x}_1, \vec{x}_2, \cdots, \vec{x}_p, \vec{u})$ denoted by $Span(\vec{x}_1, \vec{x}_2, \cdots, \vec{x}_p, \vec{u})$ and $(\vec{x}_1, \vec{x}_2, \cdots, \vec{x}_p) \perp \vec{u}$, which meet the vector algorithms. In many studies, the variables $\vec{y}$ and $\vec{x}_i$ are standardized, which is more convenient for explaining the partial correlation coefficient and $R^2$ (Thomas and O'quigley (1993)). The same was not done in this study.

In the first place, we considered a model with one explanatory variable:

$$Y = \beta_0 + \beta X_1 + u. \tag{10}$$

We also handled the sample with centralization. In geometry, its form is $\vec{y} = \beta \vec{x}_1 + \vec{u}$. Vector $\vec{y}$ is parallel to vector $\vec{Y}$. Since vector $\vec{y}$ and $\vec{x}_1$ intersect at the origin, the geometry of the univariate regression model is as follows:

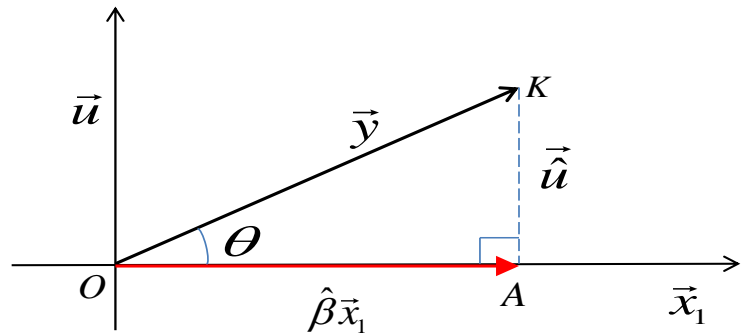

Figure 1. Geometry of the univariate linear regression model

$\hat{\vec{y}} = \hat{\beta} \vec{x}_1$ is the projection of vector $\vec{y}$ on vector $\vec{x}_1$. The cosine of angle $\theta$, which is between $\vec{y}$ and $\vec{x}_1$, is the correlation coefficient of the two variables. The error term $\vec{u}$ is not observable, which should be understood as the effect of the other variables on $y$ after removing the effect of $x_1$, and it is orthogonal to the vector $\vec{x}_1$. Vector $\vec{y}$, $\hat{\beta} \vec{x}_1$, and the error term $\vec{u}$ satisfy the Pythagorean theorem, that is $\|\vec{y}\|^2 = \|\hat{\beta} \vec{x}_1\|^2 + \|\vec{u}\|^2$. The goodness of fit is $R^2 = \|\beta\vec{x}_1\|^2 / \|\vec{y}\|^2$, $1 - R^2 = \|\vec{u}\|^2 / \|\vec{y}\|^2$, and $\|\vec{u}\| = \sqrt{1-R^2}\|\vec{y}\|$, where $\|\cdot\|^2$ is the Euclidean Norm.

We further analyzed the confidence intervals of $\hat{\beta}$: $\text{var}(\hat{\beta}) = \sigma^2 / \sum x_{1i}^2$, $se(\hat{\beta}) = \sqrt{\sigma^2 / \sum x_{1(i)}^2}$, and $\sigma^2 = \|\vec{u}\|^2 / (n-1)$, and the intercept was not considered here, so the degree of freedom $df = 1$. $|\overrightarrow{AK}| = \|\vec{u}\| = \sqrt{(n-1)\sigma^2}$ and $|\overrightarrow{OA}| = \|\hat{\beta}\vec{x}_1\| = \sqrt{\sum(\hat{\beta}x_{1(i)})^2}$, where $|\overrightarrow{AK}|$ is the length of $\overrightarrow{AK}$, $|\overrightarrow{OA}|$ is the length of $\overrightarrow{OA}$, and the confidence level $\alpha$. Then, the confidence interval of $\hat{\beta}$ can be expressed as:

$$\begin{aligned}\hat{\beta} \pm t_{\alpha/2} \cdot se(\hat{\beta}) &= \hat{\beta} \pm t_{\alpha/2} \cdot \sqrt{\sigma^2 / \sum x_{1(i)}^2} \\ &= \hat{\beta}\left(1 \pm t_{\alpha/2} \cdot \sqrt{\sigma^2 / \sum(\hat{\beta}x_{1(i)})^2}\right) \\ &= \hat{\beta}\left(1 \pm (t_{\alpha/2}/\sqrt{n-1}) \cdot (\sqrt{(n-1)\sigma^2}) / \sqrt{\sum(\hat{\beta}x_{1(i)})^2}\right). \\ &= \hat{\beta}\left(1 \pm (t_{\alpha/2}/\sqrt{n-1}) \cdot \|\vec{u}\| / \|\hat{\beta}\vec{x}_1\|\right) \\ &= \hat{\beta}\left(1 \pm (t_{\alpha/2}/\sqrt{n-1}) \cdot |\overrightarrow{AK}| / |\overrightarrow{OA}|\right)\end{aligned}$$

The significance of $\hat{\beta}$ depends on the value of $(t_{\alpha/2}/\sqrt{n-1}) \cdot |\overrightarrow{AK}|/|\overrightarrow{OA}|$. If $(t_{\alpha/2}/\sqrt{n-1}) \cdot |\overrightarrow{AK}|/|\overrightarrow{OA}| < 1$ is held, then $\hat{\beta}$ is statistically significant regardless of $\hat{\beta} > 0$ or $\hat{\beta} < 0$. It was not difficult to deduct that the $t$ value of the estimated parameter $\hat{\beta}$ is $t = \sqrt{n-1}|\overrightarrow{OA}|/|\overrightarrow{AK}| = \sqrt{n-1}\|\hat{\beta}\vec{x}_1\|/\|\vec{u}\|$ when $t > t_{\alpha/2}$, where $\hat{\beta}$ is significant. It is important to note that the $t$ value here does not have the sign of the estimated coefficient, which depends on two factors. One factor is the sample size $n$, and the other factor is $|\overrightarrow{OA}|/|\overrightarrow{AK}|$. When the sample size $n$ increases, if the true value $\beta$ of $\hat{\beta}$ is not zero, $|\overrightarrow{OA}| = \sqrt{\sum(\hat{\beta}x_{1(i)})^2} = \hat{\beta}\sqrt{\sum x_{1(i)}^2}$ and $|\overrightarrow{AK}| = \|\vec{u}\| = \sqrt{\sum \hat{u}^2}$ also roughly increase in the same proportion. Therefore, the $t$ value of $\hat{\beta}$ increases with the increase in the sample size $n$. Then, $\hat{\beta}$ becomes more significant. $ESS = \|\hat{\beta}\vec{x}_1\|^2$ and $RSS = \|\vec{u}\|^2$, then the $F$ statistic

that measures the significance of the entire model is $F = (n-1)\|\hat{\beta}\vec{x}_1\|^2 / \|\vec{u}\|^2$, which is also a function of the sample size and $F = t^2$ here.

In the second place, we considered a binary regression model:

$$Y = \beta_0 + \beta_1 X_1 + \beta_2 X_2 + u \tag{11}$$

The centralized processing of sample is $y = \hat{\beta}_1 x_1 + \hat{\beta}_2 x_2 + \hat{u}$, and the geometric analysis is shown in Figure 2.

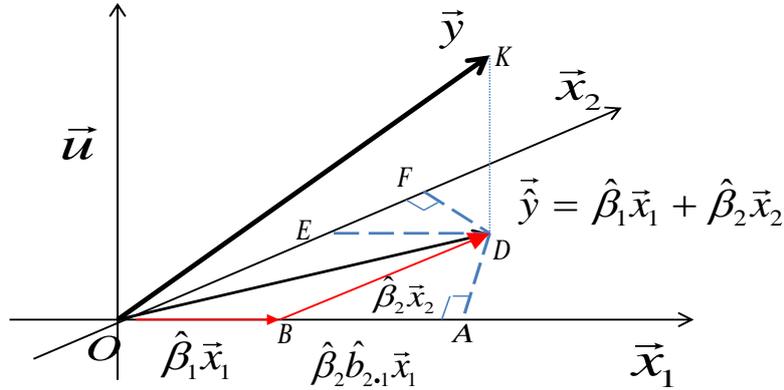

Figure 2. Geometry of the binary regression model

Obviously, vector $\hat{\vec{y}} = \hat{\beta}_1 \vec{x}_1 + \hat{\beta}_2 \vec{x}_2$ is the projection of vector $\vec{y}$ on plane $\mathrm{Span}(\vec{x}_1, \vec{x}_2)$, and $\overrightarrow{OA}$ is the projection of vector $\hat{\vec{y}} = \hat{\beta}_1 \vec{x}_1 + \hat{\beta}_2 \vec{x}_2$ on vector $\vec{x}_1$, which is consistent with the projection of $\vec{y}$ on vector $\vec{x}_1$. The projection $\overrightarrow{OA}$ is $\hat{\beta}_1 \vec{x}_1$ of the variable $y$ on $x_1$. Due to the correlation between $x_1$ and $x_2$, the projection of vector $\vec{y}$ or $\overrightarrow{OD}$ on vectors $\vec{x}_1$ and $\vec{x}_2$ are decomposed into $\hat{\beta}_1 \vec{x}_1$ and $\hat{\beta}_2 \vec{x}_2$, respectively, according to the parallelogram rule. It is clear that the projection of component $\hat{\beta}_2 \vec{x}_2$ on vector $\vec{x}_1$ is $\hat{\beta}_2 \hat{b}_{2 \cdot 1} \vec{x}_1$, so $\hat{\beta}_1 \vec{x}_1 = \hat{\beta}_1 \vec{x}_1 + \hat{\beta}_2 \hat{b}_{2 \cdot 1} \vec{x}_1$ is held. Similarly, we could analyze the explanatory variables $x_2$, $\hat{\beta}_2 \vec{x}_2 = \hat{\beta}_2 \vec{x}_2 + \hat{\beta}_1 \hat{b}_{1 \cdot 2} \vec{x}_2$. Vector $\vec{y}$ is pulled by vectors $\vec{x}_1$ and $\vec{x}_2$ and the error term $\vec{u}$.

Similarly, we further analyzed the confidence interval of the partial regression estimation coefficient. When taking $\hat{\beta}_1$ as an example: $\text{var}(\hat{\beta}_1) = \sigma^2 / \sum (x_{1i}^2 (1 - R_{12}^2))$, $R_{12}$ is the correlation coefficient of $x_1$ and $x_2$ and $\sigma^2 = \|\vec{u}\|^2 / (n-2)$, then the confidence interval of $\hat{\beta}_1$ can be expressed as:

$$\begin{aligned}
\hat{\beta}_1 \pm t_{\alpha/2} \cdot se(\hat{\beta}_1) &= \hat{\beta}_1 \pm t_{\alpha/2} \cdot \sqrt{\sigma^2 / \sum x_{1(i)}^2 (1 - R_{12}^2)} \\
&= \hat{\beta}_1 \left(1 \pm t_{\alpha/2} \cdot \sqrt{\sigma^2 / \sum (\hat{\beta}_1 x_{1(i)})^2 (1 - R_{12}^2)}\right) \\
&= \hat{\beta}_1 \left(1 \pm (t_{\alpha/2} / \sqrt{n-2}) \cdot \sqrt{(n-2)\sigma^2 / \sum (\hat{\beta}_1 x_{1(i)})^2 (1 - R_{12}^2)}\right) \\
&= \hat{\beta}_1 \left(1 \pm (t_{\alpha/2} / \sqrt{n-2}) \cdot \|\vec{u}\| / (\|\hat{\beta}_1 \vec{x}_1\| \cdot \sqrt{(1 - R_{12}^2)})\right) \\
&= \hat{\beta}_1 \left(1 \pm (t_{\alpha/2} / \sqrt{n-2}) \cdot |\overrightarrow{DK}| / (|\overrightarrow{ED}| \cdot \sqrt{(1 - R_{12}^2)})\right) \\
&= \hat{\beta}_1 \left(1 \pm (t_{\alpha/2} / \sqrt{n-2}) \cdot |\overrightarrow{DK}| / |\overrightarrow{FD}|\right)
\end{aligned},$$

where $\|\vec{u}\|$ represents the length of the error term $\vec{u}$, $|\overrightarrow{DE}| = \|\hat{\beta}_1 \vec{x}_1\|$ is the length of $\overrightarrow{OB} = \overrightarrow{DE}$, and $|\overrightarrow{FD}| = \|\hat{\beta}_1 \vec{x}_1\| \sqrt{(1 - R_{12}^2)}$ is the length of $\overrightarrow{FD}$, where the right triangle $\triangle EFD$ meets the Pythagorean theorem. Whether the estimated parameter $\hat{\beta}_1$ is significant or not depends on the size of $(t_{\alpha/2} / \sqrt{n-2}) \cdot |\overrightarrow{DK}| / |\overrightarrow{FD}|$. Even if the variable $x_1$ and variable $x_2$ are highly correlated, as long as this ratio of $|DK|/|DE|$ is reasonable, $\hat{\beta}_1$ is still significant. Therefore, with a given sample size $n$ and a confidence level $\alpha$, the significance of $\hat{\beta}_1$ not only depends on $1 - R_{12}^2$ but also on the ratio of $|\overrightarrow{DK}| / |\overrightarrow{ED}|$. It is enough that judging multiple collinearities only relies on $1 - R_{12}^2$. Analogously, we could analyze $\hat{\beta}_2$, and whether $\hat{\beta}_2$ is significant or not depends on the size of $(t_{\alpha/2} / \sqrt{n-2}) \cdot |\overrightarrow{DK}| / |\overrightarrow{AD}|$.

We also found a very interesting fact because $\triangle KFD$ is also a right triangle, which satisfies the Pythagorean theorem, namely $|\overrightarrow{FK}|^2 = |\overrightarrow{FD}|^2 + |\overrightarrow{DK}|^2$, where the three vectors separately represent the error terms of the different models. $|\overrightarrow{DK}| = \|\vec{u}\|$ represents the length of the error term $\vec{u}$ of this binary regression model, $|\overrightarrow{FD}| = \|\vec{u}_{1 \cdot 2}\|$ represents the length of the error vector of the regression model, which is the variable $\hat{\beta}_1\vec{x}_1$ on $x_2$, and $|\overrightarrow{FK}| = \|\vec{\hat{u}}\|$ represents the length of the error vector $\vec{u}$ of the univariate regression model of $y$ on $x_2$, where its expression is:

$$\begin{aligned}\vec{y} &= \hat{\beta}_1\vec{x}_1 + \hat{\beta}_2\vec{x}_2 + \vec{u} \\ &= (\hat{\beta}_1\hat{b}_{1\cdot 2}\vec{x}_2 + \vec{\hat{u}}_{1\cdot 2}) + \hat{\beta}_2\vec{x}_2 + \vec{u} \\ &= (\hat{\beta}_1\hat{b}_{1\cdot 2} + \hat{\beta}_2)\vec{x}_2 + \vec{\hat{u}}_{1\cdot 2} + \vec{u} \\ &= \hat{\tilde{\beta}}_2\vec{x}_2 + \vec{\hat{u}}\end{aligned}$$

Here, $\vec{\hat{u}} = \vec{\hat{u}}_{1\cdot 2} + \vec{u}$ and $\vec{\hat{u}}_{1\cdot 2} \perp \vec{u}$. Since $R_{12}^2 = \|\hat{\beta}_1\hat{b}_{1\cdot 2}\vec{x}_2\|^2 / \|\hat{\beta}_1\vec{x}_1\|^2$ and $R_{12}^2 = \|\hat{\beta}_2\hat{b}_{2\cdot 1}\vec{x}_1\|^2 / \|\hat{\beta}_2\vec{x}_2\|^2$, the establishment of $R_{12}^2 = \hat{b}_{1\cdot 2} * \hat{b}_{2\cdot 1}$ is clear at a glance. $1 - R_{12}^2 = \|\vec{\hat{u}}_{1\cdot 2}\|^2 / \|\hat{\beta}_1\vec{x}_1\|^2$, and the $t$ value of $\hat{\beta}_1$ can be written as the corresponding form of the vector and norm as follows:

$$\begin{aligned}t &= \sqrt{n-2} \times \frac{|\overrightarrow{ED}|}{|\overrightarrow{DK}|} \times \sqrt{1-R_{12}^2} & t &= \sqrt{n-2} \times \frac{\|\hat{\beta}_1\vec{x}_1\|}{\|\vec{\hat{u}}\|} \times \sqrt{1-R_{12}^2} \\ &= \sqrt{n-2} \times \frac{|\overrightarrow{ED}|}{|\overrightarrow{DK}|} \times \frac{|\overrightarrow{FD}|}{|\overrightarrow{ED}|} & &= \sqrt{n-2} \times \frac{\|\hat{\beta}_1\vec{x}_1\|}{\|\vec{\hat{u}}\|} \times \frac{\|\vec{\hat{u}}_{1\cdot 2}\|}{\|\hat{\beta}_1\vec{x}_1\|} \\ &= \sqrt{n-2} \times \frac{|\overrightarrow{FD}|}{|\overrightarrow{DK}|} & &= \sqrt{n-2} \times \frac{\|\vec{\hat{u}}_{1\cdot 2}\|}{\|\vec{\hat{u}}\|}\end{aligned}$$

Then, the $t$ value of $\hat{\beta}_1$ depends on the ratio of $|\overrightarrow{FD}|$ and $|\overrightarrow{DK}|$, which depends on the norm ratio of $\vec{\hat{u}}_{1\cdot 2}$ and $\vec{\hat{u}}$. In fact, $\hat{\beta}_1\vec{x}_1$ contains the indirect effect $\hat{\beta}_1\hat{b}_{1\cdot 2}\vec{x}_2$ of $x_2$, and $\vec{\hat{u}}_{1\cdot 2} = \hat{\beta}_1\vec{x}_1 - \hat{\beta}_1\hat{b}_{1\cdot 2}\vec{x}_2$ represents the net effect of $\hat{\beta}_1\vec{x}_1$ after separating the effect of $x_2$. $ESS = \|\hat{\beta}_1\vec{x}_1 + \hat{\beta}_2\vec{x}_2\|^2$ and $RSS = \|\vec{u}\|^2$, and the $F$ statistic $F = \left(\|\hat{\beta}_1\vec{x}_1 + \hat{\beta}_2\vec{x}_2\|^2 / 2\right) / \left(\|\vec{u}\|^2 / (n-2)\right)$, which is also a function of the sample size. Then, when the sample size is large enough, all the estimated parameters and the entire model statistically reach the required significance.

In the third place, we considered a ternary regression model:

$$Y = \beta_0 + \beta_1 X_1 + \beta_2 X_2 + \beta_3 X_3 + u . \tag{12}$$

The centralization of the samples is as follows. $y = \hat{\beta}_1 x_1 + \hat{\beta}_2 x_2 + \hat{\beta}_3 x_3 + \hat{u}$. The geometric analysis is shown in Figure 3. Vector $\vec{\hat{y}} = \hat{\beta}_1 \vec{x}_1 + \hat{\beta}_2 \vec{x}_2 + \hat{\beta}_3 \vec{x}_3$ is the projection of vector $\vec{y}$ on the three-dimensional space $\text{Span}(\vec{x}_1, \vec{x}_2, \vec{x}_3)$. $\hat{\beta}_1 \vec{x}_1$ is the projection of $y$ on $x_1$, namely projection $\overrightarrow{OA}$, which is divided into three partial projections: $\overrightarrow{OC}$, $\overrightarrow{CB}$, and $\overrightarrow{BA}$, respectively. These three parts represent the net effect $\hat{\beta}_1 x_1$ of $x_1$, the decomposition effect $\hat{\beta}_2 \hat{b}_{2 \cdot 1} x_1$ of $x_2$, and the decomposition effect $\hat{\beta}_3 \hat{b}_{3 \cdot 1} x_1$ of $x_3$. Obviously, $\hat{\beta}_1 x_1 = \hat{\beta}_1 x_1 + \hat{\beta}_2 \hat{b}_{2 \cdot 1} x_1 + \hat{\beta}_3 \hat{b}_{3 \cdot 1} x_1$. Therefore, in terms of force, vector $\vec{y}$ is pulled by the vectors $\vec{x}_1$, $\vec{x}_2$, and $\vec{x}_3$ in addition to the error term $\vec{u}$. The projection $\vec{\hat{y}} = \hat{\beta}_1 \vec{x}_1 + \hat{\beta}_2 \vec{x}_2 + \hat{\beta}_3 \vec{x}_3$ is composed of three vectors: $\hat{\beta}_1 \vec{x}_1$, $\hat{\beta}_2 \vec{x}_2$, and $\hat{\beta}_3 \vec{x}_3$. Also, vector $\vec{y}$ is composed of four vectors: $\hat{\beta}_1 \vec{x}_1$, $\hat{\beta}_2 \vec{x}_2$, $\hat{\beta}_3 \vec{x}_3$, and $\vec{u}$.

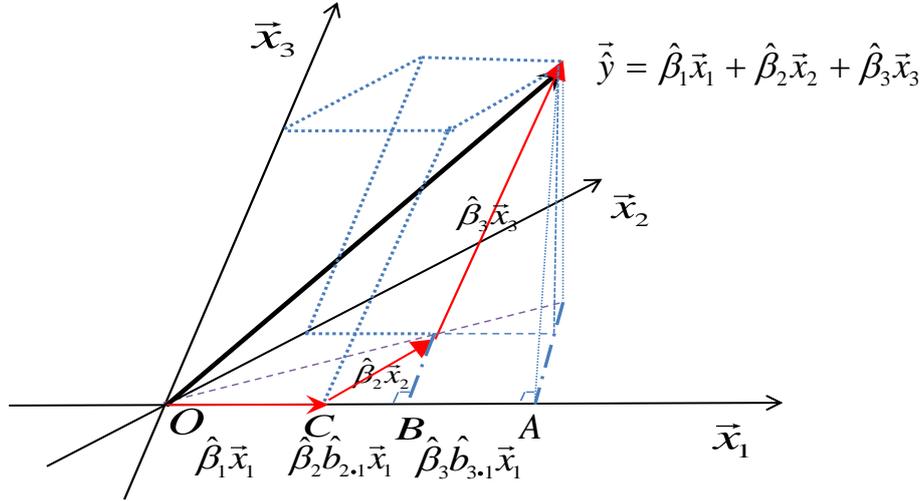

Figure 3. Geometry of the ternary regression model

According to the above analysis, the simple linear regression under the least-squares method is geometrically a vector operation, which follows the rule of parallelograms. The decomposition theorem states that unary regression coefficients are decomposed into their respective partial regression coefficients. In other words, for a multiple linear regression

model

$$Y = \beta_0 + \beta X + u, \tag{13}$$

its geometric vector form is:

$$\vec{y} = \hat{\beta}_1 \vec{x}_1 + \hat{\beta}_2 \vec{x}_2 + \cdots + \hat{\beta}_p \vec{x}_p + \vec{\hat{u}}. \tag{14}$$

Here, $\text{Span}(\vec{x}_1, \vec{x}_2, \cdots, \vec{x}_p) \perp \vec{\hat{u}}$, and the vectors $\vec{y}$, $\hat{\beta}_1 \vec{x}_1 + \hat{\beta}_2 \vec{x}_2 + \cdots + \hat{\beta}_p \vec{x}_p$, and $\vec{\hat{u}}$ satisfy the Pythagorean theorem, namely $\|\vec{y}\|^2 = \|\hat{\beta}_1 \vec{x}_1 + \hat{\beta}_2 \vec{x}_2 + \cdots + \hat{\beta}_p \vec{x}_p\|^2 + \|\vec{\hat{u}}\|^2$. If all the explanatory variables are independent of each other, then

$$\|\vec{y}\|^2 = \|\hat{\beta}_1 \vec{x}_1\|^2 + \|\hat{\beta}_2 \vec{x}_2\|^2 + \cdots + \|\hat{\beta}_p \vec{x}_p\|^2 + \|\vec{\hat{u}}\|^2.$$

For the explanatory variable $x_1$,

$$\begin{aligned}
\vec{y} &= \hat{\beta}_1 \vec{x}_1 + \hat{\beta}_2 \vec{x}_2 + \cdots + \hat{\beta}_p \vec{x}_p + \vec{\hat{u}} \\
&= \hat{\beta}_1 \vec{x}_1 + (\hat{\beta}_2 \vec{x}_2 + \cdots + \hat{\beta}_p \vec{x}_p) + \vec{\hat{u}} \\
&= \hat{\beta}_1 \vec{x}_1 + (\hat{\beta}_2 \hat{b}_{2\cdot 1} \vec{x}_1 + \cdots + \hat{\beta}_p \hat{b}_{p\cdot 1} \vec{x}_1 + \vec{\hat{u}}_{2\cdots p \cdot 1}) + \vec{\hat{u}} \\
&= (\hat{\beta}_1 + \hat{\beta}_2 \hat{b}_{2\cdot 1} + \cdots + \hat{\beta}_p \hat{b}_{p\cdot 1}) \vec{x}_1 + \vec{\hat{u}}_{2\cdots p \cdot 1} + \vec{\hat{u}} \\
&= \hat{\hat{\beta}}_1 \vec{x}_1 + \vec{\hat{\hat{u}}}
\end{aligned} \tag{15}$$

Here, $\vec{\hat{u}}_{2\cdots p \cdot 1}$ represents the sum of the error terms of the remaining single variable $x_i (i \neq 1)$ on $x_1$, including the estimated coefficients, such as $\hat{\beta}_2 \vec{x}_2$, which can also be interpreted as the error terms of $\hat{\beta}_2 \vec{x}_2 + \cdots + \hat{\beta}_p \vec{x}_p$ as a whole on $x_1$. It can also be considered that $\vec{\hat{u}}_{2\cdots p \cdot 1} = \vec{\hat{u}}_{2\cdot 1} + \vec{\hat{u}}_{3\cdot 1} + \cdots + \vec{\hat{u}}_{p\cdot 1}$, $\vec{\hat{\hat{u}}} = \vec{\hat{u}}_{2\cdots p \cdot 1} + \vec{\hat{u}}$, and $\vec{\hat{u}}_{2\cdots p \cdot 1} \perp \vec{\hat{u}}$. Equation (15) also reflects the unity of the univariate regression model and multiple regression model.

If studying the significance of $\hat{\beta}_1$, you need to remove the indirect effect of the remaining explanatory variables in $\hat{\beta}_1 x_1$, that is, to perform regression on the remaining explanatory variables as shown below.

$$\begin{aligned}
\vec{y} &= \hat{\beta}_1 \vec{x}_1 + \hat{\beta}_2 \vec{x}_2 + \cdots + \hat{\beta}_p \vec{x}_p + \vec{\hat{u}} \\
&= (\hat{\beta}_1 \hat{b}_{12} \vec{x}_2 + \cdots + \hat{\beta}_1 \hat{b}_{1p} \vec{x}_p + \vec{\hat{u}}_{1\cdot 2\cdots p}) + \hat{\beta}_2 \vec{x}_2 + \cdots + \hat{\beta}_p \vec{x}_p + \vec{\hat{u}}, \\
&= (\hat{\beta}_1 \hat{b}_{12} + \hat{\beta}_2) \vec{x}_2 + \cdots + (\hat{\beta}_1 \hat{b}_{1p} + \hat{\beta}_p) \vec{x}_p + \vec{\hat{u}}_{1\cdot 2\cdots p} + \vec{\hat{u}}
\end{aligned} \tag{16}$$

where $\hat{b}_{1i}$ represents the partial regression coefficient of the variable $x_1$ as the explained variable on the remaining explanatory variables $x_i$, namely $\vec{x}_1 = \hat{b}_{12}\vec{x}_2 + \cdots + \hat{b}_{1p}\vec{x}_p + \vec{v}_{1\cdot 2\ldots p}$. Obviously, $\vec{u}_{1\cdot 2\ldots p} = \hat{\beta}_1 \vec{v}_{1\cdot 2\ldots p}$, $\vec{v}_{1\cdot 2\ldots p} \perp \vec{u}$, and $\vec{u}_{1\cdot 2\ldots p} \perp \vec{u}$, so:

$$\vec{u}_{1\cdot 2\ldots p} + \vec{u} = \hat{\beta}_1 \vec{v}_{1\cdot 2\ldots p} + \vec{u}. \tag{17}$$

It is also obvious from equation (16) that $\vec{u}_{1\cdot 2\ldots p} + \vec{u}$ is the regression residual of the dependent variable $y$ on variables $(x_2, x_3, \cdots, x_p)$ and that $\vec{v}_{1\cdot 2\ldots p}$ is the regression residual of the dependent variable $x_1$ on variables $(x_2, x_3, \cdots, x_p)$. The conclusion here is the famous Frisch-Waugh-Lovell theorem (FWL theorem, Frisch and Waugh (1933), Lovell (1963)), which is mainly a form of decomposition theorem. Then, the significance of $\hat{\beta}_1$ in equation (16) depends on the norm ratio of vectors $\vec{u}_{1\cdot 2\ldots p}$ and $\vec{u}$. Also, $\vec{u}_{1\cdot 2\ldots p} = \hat{\beta}_1 \vec{x}_1 - (\hat{\beta}_1 \hat{b}_{12}\vec{x}_2 + \cdots + \hat{\beta}_1 \hat{b}_{1p}\vec{x}_p)$ is included in $\hat{\beta}_1 x_1$, which is the direct effect generated by $x_1$ after removing the indirect effect of the remaining independent variable $x_i (i \neq 1)$. In essence, it is to examine the significance of the estimated parameter in equation (17), which is to convert the significance of the partial regression coefficients in equation (16) into the significance of the unary regression coefficient in equation (17). Therefore, the conclusion of the FWL theorem is not complete, and not only do $\hat{\beta}_1$ and $\vec{u}$ not change, but neither does the $t$ value of $\hat{\beta}_1$. Calculate the $t$ value of $\hat{\beta}_1$ directly as follows: $t = \sqrt{n-p}\,\|\vec{u}_{1\cdot 2\ldots p}\|/\|\vec{u}\|$. If $t > t_{\alpha/2}$, then $\hat{\beta}_1$ is significant. When the sample size $n$ is small, the $t$ value becomes relatively small, so the estimated parameters are less statistically significant, especially when the number of samples is exactly equal to the number of estimated parameters, then the $t$ value is zero. In addition, $ESS = \|\hat{\beta}_1\vec{x}_1 + \hat{\beta}_2\vec{x}_2 + \cdots + \hat{\beta}_p\vec{x}_p\|^2$, $RSS = \|\vec{u}\|^2$, and the $F$ statistic that measures the significance of the entire model is expressed as: $F = \left(\|\hat{\beta}_1\vec{x}_1 + \hat{\beta}_2\vec{x}_2 + \cdots + \hat{\beta}_p\vec{x}_p\|^2 / p\right)/\left(\|\vec{u}\|^2/(n-p)\right)$, which is also a function of the sample

size. From the calculation formula of the $t$ value and $F$ statistic, it is not difficult to see why the $F$ statistic is significant and some $t$ values are not. Here, we might as well indulge ourselves if we copy the sample intact once and expand the sample size to twice the original size so we get no changes in the estimated parameters and goodness of fit. However, the $t$ values of the estimated parameter and the $F$ statistic of the model significantly increase. You can copy the sample multiple times until all the parameters are significant. Although this is a joke, it does perform so in terms of data. In essence, this is a cheating behavior, and you can also get this conclusion based on $t = \sqrt{2n-p}(2\|\vec{u}_{1\cdot2\ldots p}\|)\big/(2\|\vec{u}\|) = \sqrt{2n-p}\|\vec{u}_{1\cdot2\ldots p}\|\big/\|\vec{u}\|$.

Therefore, no matter whether the explanatory and explained variables are fixed or random, one cannot destroy the correlation structure between them. Especially, in a predicted environment, structural change testing is necessary. In almost all econometrics, the partial regression coefficient was interpreted as follows. When other variables are fixed or remain unchanged, it is the direct or net effect of the explanatory variables on the explained variables. However, in fact, the correlation between the explanatory variables was stripped away. Nevertheless, it is paradoxical that we desperately consider the correlations between the variance and significance of the partial regression coefficients. If the other variables remain unchanged, their correlation coefficient should be zero at this time, so it is recommended to measure the significance of the partial regression coefficient, which should be defined as $t^* = \sqrt{n-p}\|\hat{\beta}_1 \vec{x}_1\|\big/\|\vec{u}\|$. Obviously, $t^* \geq t$ only when the explanatory variable is not related to other explanatory variables, so $t^* = t$ is true. We vectored the variables and determined the direction, and the explanatory and explained variables were not only related but also followed the mechanics theory in physics. Some explanatory variables had a significant effect on the explained variables, although it is very difficult to explain, which can be understood as a representation of the force acting in a certain direction of the explained variable. We failed to find the original force in this direction although it numerically behaved, so we should respect the logic of things. It is similar to the instrument variable, which is basically in the same direction as the explanatory variable it replaces, except that the instrument variable is more stringent, which requires orthogonality with other explanatory variables.

## 4. Multicollinearity

The multicollinearity phenomenon is currently discussed in econometrics. One of phenomenon is that the entire regression model is quite significant and that few explanatory variables are significant. According to the above analysis, this basically led to the lack of samples due to the correlation of explanatory variables. Another phenomenon is that the signs of the partial regression coefficients of some independent variables do not match the actual situation. Simply put, there is a sign deviation between the unary regression coefficient and the partial regression coefficient, and we call it the sign expectation deviation, which refers to the phenomenon that people only intuitively observe the sign of the unary regression coefficient, where the sign of the partial regression coefficient is opposite to it. In this paper, the sign expectation deviation can be attributed to two types of reasons. One is the population structure of the explained and explanatory variables, which we call the problem of the variables structure, and the other is the problem of sample selection.

First of all, the problem of the variables structure means that some explanatory variables are very significant when the sample size is sufficient but the sign is opposite to the expected one. We used a binary linear regression as an example to graphically illustrate this problem, as shown in Figure 4.

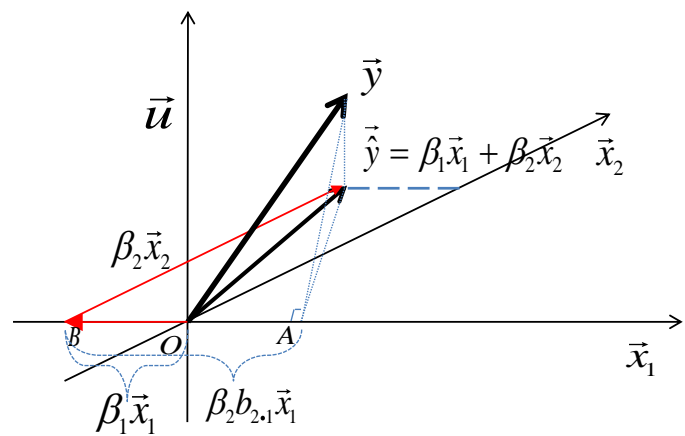

Figure 4. Geometric diagram of multicollinearity in the population structure

As shown in Figure 4, even if the correlation between $x_1$ and $x_2$ is low, the projection $\hat{y}$

of $y$ on the plane $\text{Span}(x_1, x_2)$ is on the same side of $x_1$ and $x_2$ although the projection $\overrightarrow{OA}$ of $\hat{y} = \beta_1 x_1 + \beta_2 x_2$ on $x_1$ is positive. In other words, the univariate regression coefficient of $y$ on $x_1$ is positive, and the structure of the variables $x_1$ and $x_2$ relative to $y$ is unreasonable, which makes the direct effect $\beta_1 x_1$ of $x_1$ negative, so multicollinearity occurs. In general, due to our cognitive limitations, we only intuitively observed the univariate symbols of the explained and explanatory variables and could not deeply perceive the multivariate structural relationship between them. If we had introduced polynomial terms, lag time terms, or more explanatory variables of the same time trend in the regression model, the variables structural problem could have been more common. If the model specification is not correct, adding redundant variables may lead to sign expectation deviation according to the decomposition theorem. The same is true for missing important variables, which is related to in econometrics as model specification errors. Even if it is assumed that the theoretical model of economics is correct, there is no guarantee that there is no sign deviation, as most of the theoretical models of economics do not presciently tell us that the structure of such variables leads to sign expectation deviation. If there are structural problems with the explained and explanatory variables in the population, then the increase in the sample size obviously cannot effectively reduce multicollinearity but can only improve the statistical properties of the used parameters. The correlation between explanatory variables is only a necessary condition leading to multicollinearity but not a sufficient condition. Even if explanatory variables are highly related, if the structure of the explained and explanatory variables is reasonable, multicollinearity does not occur. Conversely, even if there is a low correlation between explanatory variables, multicollinearity also occurs even if the explanatory variables structure is unreasonable. Of course, when explanatory variables are independent of each other, multicollinearity does not occur. It is a serious problem that a few classic textbooks believe that the correlation of explanatory variables may be a sufficient condition for multicollinearity rather than a necessary condition, such as Gujarati and Porter (2008) [5rd,p338]. At present, there are many methods for testing multicollinearity, such as characteristic roots and VIFs. However, they only consider explanatory variables. Therefore, these testing methods cannot be totally accepted without considering the structure of the explained and explanatory variables. In practical applications, ordinary researchers consider the correlation of explanatory variables when they have problems in establishing regression models. They believe that the correlation of explanatory variables is the unique reason for multicollinearity, which is a misunderstanding. This is because it is usually not observed that there is no multicollinearity in some models with a high correlation of explanatory variables.

Second, another reason is the sample selection problem. In large samples, the estimated parameters get closer to the true values, so the estimated parameters ultimately reflect the true structure of the explained and explanatory variables. In real economic activities, the sample size that can be obtained is limited. If the sample selection is insufficient, some important independent variables become very significant in one-variable regression, and their partial regression coefficients are not significant, even with the sign expectation deviation problem. Under the structure of $x_1$ and $x_2$, when the explained and explanatory variables did not have the problems of the above-mentioned population structure, as shown in Figure 5, (1) we discussed the cases in which the important explanatory variables were not significant and observed the $t$ value of the partial regression coefficient $\hat{\beta}_1$: $t = \sqrt{n-2} \|\vec{u}_{1\cdot 2}\| / \|\vec{u}\|$, where $\|\vec{u}_{1\cdot 2}\| = \|\hat{\beta}_1 x_1\| \sqrt{1-R_{12}^2}$, which is proportional to the absolute value of $\hat{\beta}_1$ and is approximately proportional to the sample size $n$. If the correlation between $x_1$ and $x_2$ is higher, the $t$ value of $\hat{\beta}_1$ is significantly reduced, so more samples are needed to make up for the $t$ value. At a given sample size, the smaller the absolute value of $\hat{\beta}_1$, the smaller the $t$ value and the less significant the estimated parameter. (2) We discussed the expected deviation of the symbols and observed the variance $\mathrm{var}(\hat{\beta}_1)$ of $\hat{\beta}_1$: $\mathrm{var}(\hat{\beta}_1) = \sigma^2 / \sum x_{1(i)}^2 (1-R_{12}^2)$, which is inversely proportional to the sample size and has nothing to do with the value of $\hat{\beta}_1$. Assuming the true parameter $\beta_1 > 0$, we had a higher probability that the estimated parameter $\hat{\beta}_1$ of the sample is less than zero when the sample size was small. If the correlation of the explanatory variable was higher, the probability was also larger. At the same time, since $\text{var}(\hat{\beta}_1)$ has nothing to do with the value of $\hat{\beta}_1$, the smaller the absolute value of $\hat{\beta}_1$, the greater the sign expectation deviation probability.

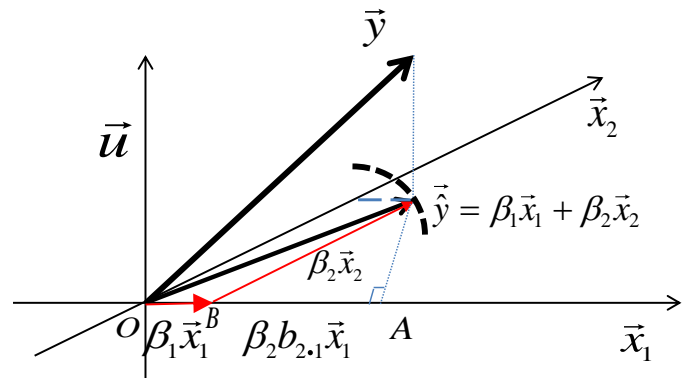

Figure 5. Geometric diagram of multicollinearity in sample selection

Conversely, due to the small sample size, the truth of multicollinearity may be masked if there is a variables structure problem, as shown above. Therefore, increasing the sample size helps in improving the statistical properties and in identifying the true structural relationship between the explained and explanatory variables. In other words, increasing the sample size is an effective method only if the true structure of variables does not have any problems.

## 5. Monte Carlo experiments

In economic statistical models, the partial regression coefficients of important variables are very significant, and the coefficient signs agree with the economic significance. Then, generally, we think that this model is excellent. As mentioned in the previous analysis, the insignificance of the partial regression coefficients of important variables is caused by insufficient samples, and the deviation of the expected sign may be an issue of the variables structure or the sample selection. Therefore, in this section, a Monte Carlo simulation was used to analyze the sign expectation deviation in multicollinearity.

Consider the following data generation process (DGP):

$$y = \beta_0 + \beta_1 x_1 + \beta_2 x_2 + u, \tag{18}$$

where $u$ follows the standard normal distribution, $x_1 \sim i.i.d.N(0,5)$, the variables $x_1$ and $x_2$ are correlated, and their correlation coefficient is $\rho$, the correlation is generated by $x_2 = \rho x_1 + \sqrt{1-\rho^2}\varepsilon$, $\varepsilon \sim N(0,5)$ and $\beta_2 = 1$, $\beta_0 = 2$. Assuming 100,000 repeat tests, let $n$ be the sample size and the correlation coefficient $\rho = 0.8, 0.5$. For the sake of simplicity, multicollinearity here means that the signs of the univariate regression coefficient and partial regression coefficient are opposite, and we defined it as the expected sign deviation.

At first, we considered the deviation of the expected sign in the population structure of variables. When the partial regression coefficient of $x_1$ in equation (18) was $\beta_1 = -0.01, -0.05, -0.1, -0.2$, it could be verified with statistical knowledge. The population univariate regression coefficients of $y$ and $x_1$ are positive when $\rho = 0.8, 0.5$. If the partial regression coefficient is negative, we believe that there is a deviation in the expected sign. In fact, we know that the sign actually does not have a deviation at this time. The statistical results are shown in Table 1.

Table 1. The deviation of the expected sign in the population structure of variables

| $n$ | $\rho$ | 0.8 | | | | 0.5 | | | |
|---|---|---|---|---|---|---|---|---|---|
| | $\beta_1$ | −0.01 | −0.05 | −0.1 | −0.2 | −0.01 | −0.05 | −0.1 | −0.2 |
| 30 | | 56.18% | 78.07% | 93.82% | 99.84% | 58.69% | 86.91% | 98.54% | 100.00% |
| 50 | | 58.20% | 84.93% | 97.93% | 100.00% | 61.71% | 93.03% | 99.78% | 100.00% |
| 100 | | 61.41% | 92.94% | 99.84% | 100.00% | 66.48% | 98.28% | 100.00% | 100.00% |

Table 1 shows that on one hand, the proportion of determining multicollinearity is increasing with the increase in the sample size. On the other hand, with a certain sample size and correlation coefficient, the smaller the value of $\beta_1$, the greater the proportion of determining multicollinearity. Therefore, increasing the sample is ineffective when there is a deviation in the expected sign in the population structure of the variables.

Second, we considered the deviation of the expected sign in the sample selection. When the partial regression coefficient was $\beta_1 = 0.01, 0.05, 0.1, 0.2$ in equation (18), it was clear that the proportion univariate regression coefficients of $y$ to $x_1$ was positive when $\rho = 0.8, 0.5$. If $\hat{\beta}_1 < 0$, there might be a deviation in the expected sign. In fact, the sign did not have a deviation at this time. The statistical results are shown in Table 2.

Table 2. Deviation of the expected sign in the sample selection

| $n$ | $\rho$ | 0.8 | | | | 0.5 | | | |
|---|---|---|---|---|---|---|---|---|---|
| | $\beta_1$ | 0.01 | 0.05 | 0.1 | 0.2 | 0.01 | 0.05 | 0.1 | 0.2 |
| 30 | | 43.78% | 21.60% | 6.13% | 0.18% | 40.86% | 13.06% | 1.57% | 0.00% |
| 50 | | 41.64% | 15.38% | 2.25% | 0.01% | 38.26% | 7.06% | 0.21% | 0.00% |
| 100 | | 38.12% | 7.12% | 0.18% | 0.00% | 33.32% | 1.77% | 0.00% | 0.00% |

Table 2 shows that under a fixed correlation coefficient and estimated parameter $\beta_1$, as the sample size increases, the proportion of the determined multicollinearity is decreased. Also, under the condition that the sample size and correlation coefficient are constant, with the increase in the value of the partial regression coefficient $\beta_1$, the proportion of multicollinearity is decreased. Moreover, with a given sample size and estimated parameters, the stronger the correlation of the independent variables, the greater the proportion of multicollinearity. It is useful to increase the samples only if there is no sign deviation in the variables structure.

The geometric analysis of multicollinearity has been verified in experimental simulations. We concluded that it is inappropriate to describe multicollinearity as a sample phenomenon, as the insignificance of the partial regression coefficients of important explanatory variables is a sample phenomenon. The phenomenon of the expected sign deviation may be a problem of the variables structure, which is our insufficient cognition, as a result, increasing the sample size cannot help. And it may also be a sample selection problem due to insufficient samples. If the sample size is increased, this phenomenon disappears. Therefore, increasing the sample size is helpful in eliminating the cause of the expected sign deviation.

# 6. Deal with multicollinearity

Many researchers are annoyed by multicollinearity and have proposed some solutions to it, such as ridge estimation, principal component analysis, elimination of variables, deletion of outliers and differences, etc. Here, we analyzed these methods one by one.

**6.1 Ridge estimation method**

The ridge estimation method was proposed by Hoerl and Kennard (1970). For a regression model with multicollinearity, the estimation coefficient of the regression model is modified to $\hat{\beta}(\lambda) = (X^T X + \lambda I)^{-1} X^T y$, where $\lambda > 0$, the purpose of which is to improve the singularity of the matrix $X^T X$. According to the above geometric analysis, the addition of $\lambda I$ destroys the parallelogram rule, which makes $\hat{\beta}(\lambda)$ biased, and $\lambda > 0$ shrinks the parallelogram inward. Since the univariate regression coefficient is positive, the method forcibly compresses the negative partial regression coefficients to positive. When $\lambda \to \infty$, then it compresses all the estimated coefficients to zero, and the value of $\lambda$ becomes very arbitrary, so there is no recognized standard. In fact, this standard may not exist in geometry.

**6.2 Principal component processing**

It is believed that multicollinearity is caused by the strong correlation between explanatory variables. The same number of mutually orthogonal variables can be generated by the principal component analysis of the original variables. You can use these conversion variables to get a new regression model without multicollinearity. This approach is equivalent to transforming an explanatory variable space into another mutually orthogonal space, which seems to be valid for multicollinearity in the population structure. However, in fact, according to the decomposition theorem, if these transformation variables are replaced by the original explanatory variables into a regression model, the parameter is still the parameter that is consistent with the original model. The principal component processing method does not differ much from the ordinary linear transformation. We used a binary regression model $y = \beta_0 + \beta_1 x_1 + \beta_2 x_2 + u$ as an example to illustrate this problem. When multicollinearity existed in the population structure of the model, we used $z_1 = x_1 + x_2$ and $z_2 = x_1 - x_2$ for the linear transformation. Then, the new model $y = b_0 + b_1 z_1 + b_2 z_2 + \varepsilon$ may not have multicollinearity. However, we replaced $z_1 = x_1 + x_2$ and $z_2 = x_1 - x_2$ into the new regression model again, and the estimated parameters still remained unchanged. Also, multicollinearity was not really eliminated, and it was just a psychological comfort for the model builder. There is not much difference between the principal component analysis, factor analysis, partial least squares, path analysis, and ordinary linear transformation. As long as it is a linear transformation, its essence is not changed according to the decomposition theorem. Of course, if it is a non-linear transformation, such as a ratio transformation, and there may not be multicollinearity in the new model, but it is no longer the structure of the original linear model.

**6.3 Elimination of variables and outliers**

To eliminate multicollinearity, one of the simplest ways is to eliminate one of the collinear variables, and the stepwise regression method is probably most frequently used. Based on the above multicollinearity analysis, if there is multicollinearity in the variables structure in a model, we generally think that there are more redundant independent variables, so it seems desirable to eliminate some variables, but sometimes we may omit some important variables, so eliminating variables not only affects the fitting effect of the model but can also possibly remove some important variables, as with a given sample size, people pay more attention to the $p$ value. If multicollinearity is caused by the sample selection, it is obvious that excluding any variables will not only lead to the loss of the important explanatory variables but may also affect the model's fitting effect. The removal of the abnormal points of multicollinearity means that a certain point is considered as an abnormal point of data after diagnosis, which is removed from the sample in order to improve the degree of multicollinearity. This method is effective under the multicollinearity caused by sample selection, but there is almost no effective method regarding the multicollinearity caused by the variables structure, so the prerequisite is that the essence of multicollinearity must be clarified.

**6.4 Difference method**

The difference method entails the transformation of original models into differential models. Some believe that it can effectively eliminate multicollinearity. Here, we need to introduce the difference structure invariance theorem to explain the role of the difference method. Before proving this theorem, we need to provide Lemma 2.

**Lemma 2.** In large samples, for a univariate regression model $y = \hat{\beta}_0 + \hat{\beta}x + u$, the regression parameter is $\hat{\beta}$, and the regression parameters in its differential model $\Delta y = \tilde{\beta}\Delta x + \Delta u$ is $\tilde{\beta}$. So,

$$P\lim(\hat{\beta}) = P\lim(\tilde{\beta}) = \beta. \tag{19}$$

This is because the variables $x$ and $u$ are independent of each other, so the variables $\Delta x$ and $\Delta u$ must also be independent of each other. Then, it can be proven according to the regression parameters form. The proof process is omitted here. In a small sample, $\hat{\beta} \neq \tilde{\beta}$, and when the samples gradually increases to infinity, they all tend to be true $\beta$, and $VAR(\hat{\beta}) < VAR(\tilde{\beta})$. Although $\hat{\beta}$ and $\tilde{\beta}$ are unbiased estimates, both are samples of the population parameter $\beta$, but $\hat{\beta}$ is the optimal unbiased estimate (BLUE). With Lemma 2, it is easy to get the difference structure invariance theorem as follows.

**Proposition 1.** Based on the least-squares method, if the parameter structure of the original multiple linear model is $\hat{\beta} = [\hat{\beta}_1, \hat{\beta}_2, \hat{\beta}_3, \cdots, \hat{\beta}_p]'$, the parameter structure of the regression model obtained by the difference model is $\tilde{\beta} = [\tilde{\beta}_1, \tilde{\beta}_2, \tilde{\beta}_3, \cdots, \tilde{\beta}_p]'$. Then,

$$P\lim(\hat{\beta}) = P\lim(\tilde{\beta}) = \beta. \tag{20}$$

According to the decomposition theorem, the partial parameter regression structure can be written as:

$$\begin{pmatrix} \hat{\beta}_1 \\ \hat{\beta}_2 \\ \vdots \\ \hat{\beta}_p \end{pmatrix} = B^{-1} \begin{pmatrix} \hat{\hat{\beta}}_1 \\ \hat{\hat{\beta}}_2 \\ \vdots \\ \hat{\hat{\beta}}_p \end{pmatrix} = \begin{pmatrix} \hat{b}_{1 \cdot 1} & \hat{b}_{2 \cdot 1} & \cdots & \hat{b}_{p \cdot 1} \\ \hat{b}_{1 \cdot 2} & \hat{b}_{2 \cdot 2} & \cdots & \hat{b}_{p \cdot 2} \\ \vdots & \vdots & \cdots & \vdots \\ \hat{b}_{1 \cdot p} & \hat{b}_{2 \cdot p} & \cdots & \hat{b}_{p \cdot p} \end{pmatrix}^{-1} \begin{pmatrix} \hat{\hat{\beta}}_1 \\ \hat{\hat{\beta}}_2 \\ \vdots \\ \hat{\hat{\beta}}_p \end{pmatrix}.$$

Obviously, all the right elements of equation (21) are the estimated parameters of the one-variable regression model. According to Lemma 2, for the original model and difference model, with the large sample, both the right sides were consistent, so their partial regression parameters were also equal. That is to say, by comparing the original model with the difference model in a large sample, the structures of the univariate regression parameters of the explained variables on the explanatory variables and the partial regression parameter structure did not change. With a small sample, the parameter structures of the original model and difference model were just a sampling of the population parameter structure, but the parameter of the original model maintained the BLUE property.

The difference model does not add new data and only transforms original data without adding new information, which does not help in estimating any parameters. However, it is still useful in detecting multicollinearity. According to the parameter structure invariance theorem for large samples, the parameter structure of a model remains unchanged after the original data is differentiated. Therefore, the difference method helps in identifying the cause of multicollinearity. Similar to the role of the difference method in non-stationary time sequences, due to the lack of samples in the original sequence, there may be a linear correlation between the variables that are originally irrelevant, so the non-stationary sequence needs to be changed into a stationary sequence through the difference method. The difference method is used to identify the true parameters, and it is the application of the invariance theorem of the difference parameter structure. When our sample size could not be increased, because the parameter structure of the difference model and original model was a sample of the population parameter structure, the sign deviation probability between the two models was relatively small. Therefore, in practice, due to the limited number of obtained samples, considering that there may be a sign deviation in the original model, it is recommended to further try making a difference model to check the sign deviation. However, it should be noted that the difference model structure is just a non-core evidence and that an accurate method still needs to increase the sample size.

In short, blind remedies can lead to worse results if applied before distinguishing the causes of multicollinearity. If there is a variables structure problem, it seems futile to solve multicollinearity without changing the variables structure of the original model. If it is a problem of the sample selection, increasing the sample size can help in solving the essence of the problem. However, with smaller sample sizes, the sign deviation problem needs to be differentiated, which is not an easy process.

## 7. Conclusion

The multicollinearity problem has been the focus of debate in academia for a long time. Through the geometric interpretation of the decomposition theorem, this paper provided two reasons for multicollinearity. One is the problem of the population structure of the explained and explanatory variables, and the other is the sample selection problem. In multicollinearity, it is a sample phenomenon that the partial regression coefficients of important explanatory variables are not significant, and it is caused due to insufficient sample size. Also, the sign deviation phenomenon may be a problem of the variables structure or a problem of the sample selection. Some of the current diagnostic methods only consider the correlation of the explanatory variables, which makes them unreliable. In fact, the high correlation of explanatory variables does not necessarily lead to multicollinearity. It can only be said that the stronger the correlation of explanatory variables, the greater the probability of multicollinearity. However, the occurrence of multicollinearity has an inevitable correlation of explanatory variables. In terms of the necessary measures for solving multicollinearity, these methods have not solved multicollinearity in essence before identifying the cause of multicollinearity. However, increasing the sample size helps us understand the cause of multicollinearity, and the difference method plays an auxiliary role.

Based on this study, we believe that if it is a problem of the variables structure, it may be a good choice to eliminate some variables, but the purpose is to predict the explained variables. Also, the elimination of variables affects the model fitting effect, and multicollinearity is not a problem. If it is a sample selection problem, deleting the outliers of data may result in a reliability estimate, which is definitely useful. According to experience, partial regression coefficients are more significant when the sample size is large. If there is a sign expected deviation, it may generally be a problem of the variables structure. Also, under a small sample size, if a sign expected deviation occurs, it may be a problem of the variables structure or a problem of the sample selection. In the presence of restrictions on increasing the samples, the difference method can be used for further testing. If the unary regression coefficients with important explanatory variables are significant, the partial regression coefficients are not significant, and that is basically a sample selection problem. When the sample size gradually increases, the partial regression coefficient gradually becomes significant.

## Acknowledgments

I thank the referee for the very helpful comments on an early version of this paper. This work was supported by Department of Education of Guangdong Province under Grant Nos.2020WTSCX261.

## Declaration of interest

## Funding

## Author Contribution